\documentstyle[prl,aps,twocolumn]{revtex}

\begin{document}
\input epsf
\twocolumn[\hsize\textwidth\columnwidth\hsize\csname
@twocolumnfalse\endcsname

\title{Direct link between Coulomb blockade and shot noise in a quantum
coherent structure}
\author{A. Levy Yeyati$^{1}$, A. Martin-Rodero$^{1}$, D. Esteve$^{2}$ and C.
Urbina$^{2}$}
\address{$^{1}$Departamento de F\'\i sica Te\'orica de la Materia 
Condensada C-V. \\
Facultad de Ciencias, \\
Universidad Aut\'onoma de Madrid, E-28049 Madrid, Spain \\
$^{2}$Service de Physique de l'Etat Condens\'{e}, Commissariat \`{a} \\
l'Energie Atomique, Saclay, F-91191 Gif-sur-Yvette Cedex, France}
\maketitle

\begin{abstract}
We analyze the current-voltage characteristic of a quantum conduction
channel coupled to an electromagnetic environment of arbitrary 
frequency-dependent impedance. In the weak blockade regime  
the correction to the ohmic behavior is directly related to the channel 
current fluctuations
vanishing at perfect transmission in the same way as shot noise. 
This relation can be generalized to describe the environmental Coulomb
blockade in a generic mesoscopic conductor coupled to an
external impedance, as the response of the latter to the current
fluctuations in the former.
\end{abstract}

\vskip2pc]

\narrowtext

The way in which quantum mechanics affect the laws ruling electrical
circuits is presently understood only for some elementary situations.
For instance, in the case of a quantum coherent nanostructure connecting two
independent electron reservoirs with a voltage difference $V$, quantum
mechanics results in current fluctuations, the so-called shot noise,
which at low
frequency and zero temperature have a spectrum of the form 
$S=2eVG_{0}\sum \tau _{i}(1-\tau _{i})$, where the $\{\tau _{i}\}$ are
the transmissions of the conduction channels and $G_{0}=2e^{2}/h$ is the
conductance quantum \cite{noise1}. The reduction of noise with increasing
transmission as $(1-\tau _{i})$ is a consequence of the Fermi statistics in
the reservoirs, a prediction which has been tested quantitatively in
different types of nanostructures \cite{exp-noise}. Another important
consequence of quantum mechanics is that Ohm's law does not apply when two
elements are connected in series, because each element is not simply voltage
biased. The phase across each element $\phi =\frac{e}{\hbar} \int v(t)dt$,
where $v(t)$ is the voltage drop,
develops quantum fluctuations, and the electrical properties of the
series-connection cannot be inferred from the conductance of the separate
elements. More generally, the phase differences add in series, and parallel
connected branches share the same phase, but the general rules to predict the
properties of the whole circuit from those of the constitutive
elements are not known, except for macroscopic electromagnetic impedances.
When a low transmissive nanostructure, i.e. one with negligible noise 
reduction, is
connected in series with a macroscopic impedance $Z(\omega )$, the
conductance of the series circuit is suppressed at sufficiently low voltage
and temperature \cite{devoret}. This phenomenon, called environmental Coulomb
blockade, has been thoroughly investigated in small metallic tunnel
junctions \cite{junctions}. How is this phenomenon modified in the case of a
coherent structure whose transmissions $\tau _{i}$ approach unity? One might
speculate that a noiseless structure cannot be ``felt'' by the series
impedance $Z(\omega)$, and, reciprocally, should not be affected by its
presence. We show here that this naive reasoning which predicts the
restoration of Ohm's law at large transmission is correct, and more
precisely that Coulomb blockade is modified in the same way as shot noise.

In this Letter we address the simple case of a single channel 
quantum point contact with transmission $\tau $ connected in series
with a macroscopic impedance $Z(\omega )$. 
We develop a theory of the environmental Coulomb blockade which
permits to analyze the current-voltage characteristic at arbitrary
transmission for a generic frequency-dependent impedance. We show that
in the limit of low impedance $Z\ll 1/G_{0}$ the blockade is intimately
connected with the fluctuations in the current through the channel.

A single channel contact of arbitrary transmission between two 
electrodes can be modeled with a simple Hamiltonian 
resembling the usual tunneling Hamiltonian for a tunnel junction 
\cite{cuevas}. The coupling to the environment can be then introduced in the
usual way \cite{devoret}, which leads to a model Hamiltonian 
$\hat{H}=\hat{H}_{L}+\hat{H}_{R}+\hat{H}_{T}$, where $\hat{H}_{L,R}$ 
describe the uncoupled
left and right leads, characterized by flat densities of states $\rho
_{L,R}\simeq 1/\pi W$ and

\begin{equation}
\hat{H}_{T}=\sum_{\sigma }T_{0}\hat{c}_{L\sigma }^{\dagger }\hat{c}_{R\sigma
}\hat{\Lambda}_{e}+\,\,h.c.,  \label{hamiltonian}
\end{equation}
describes the transfer of an electron between the leads in terms of a
hopping element $T_0$. The translation operator 
$\hat{\Lambda}_{e}=e^{i\hat{\phi}}$ , where
$\hat{\phi}$ is the phase operator satisfying the commutation relation 
$[\hat{Q},\hat{\phi}]=ie$, takes
into account the change in the charge $\hat{Q}$ of the
environment associated with the transfer process. 
If the coupling to the environment is neglected the normal transmission of
this model is given by $\tau =4\beta /(1+\beta )^{2}$,
where $\beta =(T_{0}/W)^{2}$ \cite{cuevas}.

We are interested in calculating the current through the channel under a
constant bias voltage $V$ in the presence of the environment. The current
operator within this model is given by

\begin{equation}
\hat{I} = \frac{ie}{\hbar} \sum_{\sigma} T_0 \hat{c}^{\dagger}_{L\sigma} 
\hat{c}_{R\sigma} \hat{\Lambda}_e - \,\, h.c. ,  \label{current}
\end{equation}

In order to evaluate the mean current we use the Keldysh formalism \cite
{keldysh} which is suitable for calculating averages in a non-equilibrium
state. The mean current is formally given by

\begin{equation}
<\hat{I}(t)> = <\hat{T}_c \left[ \hat{I}_I(t) \hat{S}_c(\infty,-\infty) %
\right]> ,
\end{equation}
where $\hat{T}_c$ is the chronological ordering operator along the Keldysh
contour, $\hat{I}_I$ is the current operator in the interaction
representation and $\hat{S}_c(\infty,-\infty)$ is the corresponding
evolution operator along the closed time contour. Introducing the series
expansion of $\hat{S}_c$ in terms of $\hat{H}_T$ and applying Wick theorem
we obtain a perturbative expansion for the evaluation of the current which
can be expressed in terms of Keldysh Green functions. The lowest order
diagrams within this theory are depicted in Fig. 1a. In these diagrams we
associate a full line with an arrow to the electron propagators for the
uncoupled leads (denoted below by $g^{\alpha,\beta}_{L,R}$), crosses
indicate hopping events, and wavy lines correspond to the environment
correlators given by \cite{comment}

\begin{equation}
P^{\alpha,\beta}(t,t^{\prime}) = e^{J^{\alpha,\beta}(t,t^{\prime})} ,
\end{equation}
where $\alpha,\beta \equiv +,-$ indicate the branch on the Keldysh contour
for the two time arguments, and $J^{\alpha,\beta}$ are the phase correlation
functions 
\[
J^{\alpha,\beta}(t,t^{\prime}) = <\hat{T}_c \left[\hat{\phi}(t_{\alpha}) 
\hat{\phi}(t^{\prime}_{\beta}) \right]> - <\hat{\phi}^2> . 
\]

As usual, we assume that the modes in the environment (or photon states) are
populated according to the equilibrium distribution at a given temperature.
In this case $J^{\alpha,\beta}(t,t^{\prime}) =
J^{\alpha,\beta}(t-t^{\prime}) $.

The evaluation of the complete perturbative series is a formidable task. One
can, however, obtain useful results in the limit of weak impedance $Z \ll
1/G_0$. In this limit $P^{\alpha,\beta}$ can be approximated by $1 +
J^{\alpha,\beta}$. To the first order in $J^{\alpha,\beta}$ one obtains the
family of diagrams depicted in Fig 1b, where the phase correlation functions
are represented by a dashed line. These diagrams correspond to single
``photon'' processes and can be associated into four groups depending on the
types of hopping event (left to right or right to left) connected by the
wavy line. By introducing renormalized electron propagators and renormalized
hopping amplitudes each one of these groups gives rise to a diagram like the
one depicted in Fig. 1c. The remaining part of the calculation relies in
obtaining the expression of these diagrams in terms of Keldysh Green
functions.

A further simplification of the calculation is obtained by considering the
wide band limit, i.e. to assume that the electron band width is much larger
than all other relevant energy scales involved in the problem. Within this
approximation the propagators for the isolated leads are given by

\begin{eqnarray}
\hat{g}_{L,R}(\omega) &=& \left( 
\begin{array}{cc}
g^{++}_{L,R}(\omega) & g^{+-}_{L,R}(\omega) \\ 
g^{-+}_{L,R}(\omega) & g^{--}_{L,R}(\omega)
\end{array}
\right) = \nonumber \\
& &\frac{i}{W} \left( 
\begin{array}{cc}
2 f_{L,R}(\omega) -1 & 2 f_{L,R}(\omega) \\ 
2(f_{L,R}(\omega)-1) & 2f_{L,R}(\omega) - 1
\end{array}
\right) ,
\end{eqnarray}
where $f_{L,R}(\omega)$ are the Fermi distribution functions on the left and
right lead respectively.

Another basic ingredient in the calculation is the convolution of electron
propagators with phase correlations $\delta g^{\alpha,\beta}_{L,R}(\omega) =
\int d\omega^{\prime} J^{\alpha,\beta}(\omega^{\prime})
g^{\alpha,\beta}_{L,R}(\omega+\omega^{\prime})$. In the wide band limit
these can be approximated as

\begin{equation}
\delta \hat{g}_{L,R}(\omega) = \frac{i}{W} \left( 
\begin{array}{cc}
f_{L,R}^+(\omega) + f_{L,R}^-(\omega) & 2 f_{L,R}^+(\omega) \\ 
2 f_{L,R}^- & f_{L,R}^+(\omega) + f_{L,R}^-(\omega)
\end{array}
\right) 
\end{equation}
where $f_{L,R}^{\pm}(\omega) = \int d\omega^{\prime} J(\omega^{\prime})
f_{L,R}(\omega\pm\omega^{\prime})$, $J(\omega)=J^{+-}(\omega)$ being the
Fourier transform of the phase correlation function.

For an energy independent transmission coefficient one then obtains the
following expression for the correction to the current induced by the
environment:

\begin{eqnarray}
\delta I(V) &=& \frac{e}{h} \tau (1 - \tau) \int d\omega \left[f_R(\omega)
\left(f^-_L(\omega) - f^+_L(\omega)\right) \right. \nonumber \\
&& \left. - f_L(\omega) \left(f^-_R(\omega)
- f^+_R(\omega)\right) \right]  \nonumber \\
&& + \frac{e}{h} \tau^2 \int d\omega \left[f_L(\omega) \left(f^-_L(\omega) 
- f^+_L(\omega)\right) \right. \nonumber \\
&& \left. - f_R(\omega) \left(f^-_R(\omega) -
f^+_R(\omega)\right) \right].  \label{final-expression}
\end{eqnarray}

By analyzing the Fermi factors, this expression can be decomposed as 
$\delta I = \delta I^{\rightarrow}
- \delta I^{\leftarrow}$, where $\delta I^{\rightarrow}$ and
$\delta I^{\leftarrow}$  
correspond to currents flowing in the two opposite directions.  
Both terms can be interpreted as arising from
the coupling between the 
contact current fluctuations and the phase fluctuations due to the
finite impedance of the environment. In fact, as described below, 
this expression can be directly related to the current fluctuations. 
In the absence of environment the noise spectrum 
in a single channel contact is given by \cite{Khlus}

\begin{eqnarray}
S(V,\Omega) &=& \frac{2e^2}{h} \tau (1 - \tau) \int d\omega \left[f_R(\omega)
\left(1 - f_L(\omega+\Omega)\right) \right. \nonumber \\
&& \left. + f_L(\omega) \left(1 - f_R(\omega+\Omega)\right) \right]  \nonumber \\
&& + \frac{2e^2}{h} \tau^2 \int d\omega \left[f_L(\omega) \left(1 -
f_L(\omega+\Omega)\right) \right. \nonumber\\
&& \left. + f_R(\omega) \left(1 -
f_R(\omega+\Omega)\right) \right] \nonumber \\
&& + \left[\Omega \rightarrow -\Omega \right].  \label{noise-spectrum}
\end{eqnarray}

On the other hand $S(V,\Omega) = \int dt e^{i\Omega t}
\left[K(t)+K(-t)\right]$ where 
$K(t) = <\hat{I}(t)\hat{I}(0)> - <\hat{I}^2>$ is the current
correlation function. By comparing Eqs. (\ref{final-expression}) and
(\ref{noise-spectrum}) we arrive to the simple relation  

\begin{eqnarray}
e \left(\delta I^{\rightarrow} + \delta I^{\leftarrow} \right)
&=& \int dt J(t) \left[K(t) - K(-t) \right].  \label{general}
\end{eqnarray}

This expression can be considered as a generalization of the
fluctuation-dissipation theorem to the present non-equilibrium situation. As
in the low impedance regime the coupling between the contact and its
environment is of the form $\hat{I}\hat{\phi}$ we expect this result to be
valid for a generic situation with the same type of
system-environment interaction. In particular the result (\ref{general})
would apply for any mesoscopic conductor that can be modeled as a collection
of channels.

In order to understand the effects on the $I-V$ characteristic it is
instructive to consider first the case of an environment with just 
a single mode at zero temperature, for which $J(\omega )=\pi
G_{0}Z_{0}(\delta (\omega -\omega _{0})-\delta (\omega ))/2$ \cite{ingold}.
For this model the conductance exhibits a discontinuity at 
$eV=\hbar\omega _{0}$.
For $eV<\hbar\omega _{0}$ 
there is a reduction in the conductance $\delta G=-\pi
G_{0}^{2}Z_{0}\tau (1-\tau )/2$ while for $eV>\hbar\omega _{0}$, $\delta G=0$.
This simple case shows that the blockade is proportional to shot noise
and vanishes at perfect transmission.

In the general situation the environment is characterized by a complex
impedance $Z(\omega )$. The phase correlation function is related to the
impedance by the expression \cite{ingold}

\begin{equation}
J(t) = G_0 \int d\omega \frac{\mbox{Re} Z(\omega)}{\omega} \frac{e^{i\omega
t}-1}{1 - e^{-\beta \hbar \omega}} ,
\end{equation}
which at zero temperature leads to a correction in the differential
conductance given by

\begin{equation}
\frac{\delta G}{G} = - G_0 (1 - \tau) 
\int_{eV}^{\infty} d\omega \frac{\mbox{Re}%
Z(\omega)}{\omega} .  \label{ohmic}
\end{equation}

This is the same result one obtains for a tunnel junction except for
the reduction factor $(1 - \tau)$. 
In the simple but realistic case in which the impedance $Z(\omega )$
is composed by the resistance $R$ of the leads embedding the contact in
parallel with the capacitance $C$ of the contact itself, 
$Z(\omega)=R/(1+i\omega RC)$, 
and the integral in (\ref{ohmic}) yields

\begin{equation}
\frac{\delta G}{G} = - G_0 R (1-\tau) \ln\sqrt{1 + 
\left(\frac{\hbar\omega_R}{eV} \right)^2 } ,
\label{ohmic2}
\end{equation}
where $\omega_R = 1/RC$. For finite temperature $\delta G$ can be evaluated
numerically from Eq. (\ref{final-expression}). Fig. 2 shows the evolution of
the correction to the differential conductance with temperature. For an
energy independent transmission the temperature dependence is the same as
for a tunnel junction except for a a global factor $(1-\tau)$. For
increasing temperatures the dip in the conductance at zero bias is
progressively washed out.

The multichannel extension of these results is straightforward. One
can for instance analyze the case of a diffusive conductor by
replacing the reduction factor $(1-\tau)$ by its average over many
channels (which in the diffusive case is known to be $1/3$) and $R$ by the
resistance of the conductor itself \cite{zaikin}. 
This analysis leads to quantitative
agreement with recent experimental results \cite{weber}.

As a final remark we would like to stress that the results of the present
theory can be thoroughly tested experimentally using atomic contacts that 
can be produced by scanning tunneling microscope or break junction techniques 
\cite{ruitenbeek}. 
These contacts accommodate a small number of channels,
and the ensemble of the transmissions $\{\tau _{i}\}$ can be
determined experimentally and varied over a wide range including 
the $\tau \rightarrow 1$ limit \cite{scheer}.
Moreover, the impedance of
the environment embedding such contacts can be tuned within a desired range
using nanolithography \cite{goffman}. Experimental work along these lines is
currently under progress.

In conclusion we have presented a theoretical analysis of the environmental
Coulomb blockade in coherent nanostructures. We have considered the
weak blockade regime and showed that the corrections in the current-voltage
characteristic can be related to the structure current fluctuations.
The blockade vanishes at perfect transmission as $\tau (1-\tau )$, in the
same way as shot noise. The temperature dependence of the conductance 
is similar to the one observed in ultrasmall tunnel junctions. The present
calculation provides a first step towards the understanding of Coulomb
blockade effects in coherent nanostructures. Although derived for a
particular model it has been argued that the expression (\ref{general}) is
more general and valid for a generic mesoscopic conductor coupled to an 
arbitrary external impedance. The strong blockade limit could be
addressed through the analysis of multiple photon processes along the 
lines suggested in this work.

\acknowledgements
We acknowledge fruitful discussions with R. Cron, J.C. Cuevas,  M.
Devoret, M. Goffman and P. Joyez. We also thank A.D. Zaikin for calling
to our attention the experimental results of Ref. \cite{weber}.
This work was partially supported by BNM and MAE through PICASSO and
the Spanish CICyT under contract PB97-0044.

\begin{figure}[tbp]
\begin{center}
\leavevmode
\epsfysize=10cm
\epsfbox{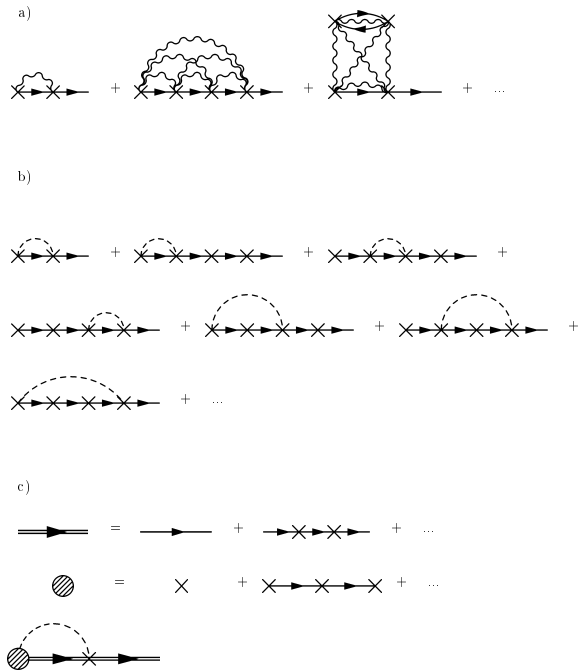}
\end{center}
\caption{Diagrammatic expansion of the current in the Keldysh formalism: a)
Unlabeled lowest order diagrams in $\hat{H}_T$. Full lines with an arrow
indicate electron propagators, crosses indicate hopping events and wavy
lines correspond to environment correlators. b) Single photon processes up
to second order in $\hat{H}_T$. Dashed lines indicate phase correlation
functions. c) Renormalized diagrams arising from the addition of single
photon processes up to infinite order in the hopping. Double full lines with
an arrow indicate dressed electron propagators and shaded circles indicate
dressed hopping amplitudes.}
\end{figure}

\begin{figure}[tbp]
\begin{center}
\leavevmode
\epsfysize=5cm
\epsfbox{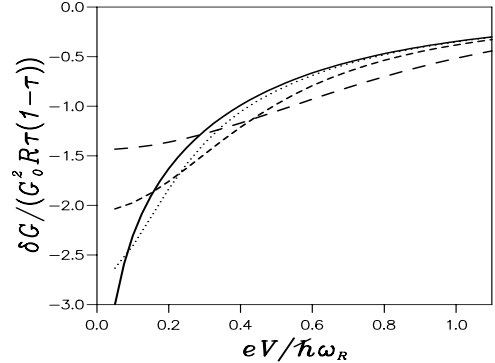}
\end{center}
\caption{Correction to the conductance of a single conduction channel
with transmission $\tau$ due to an ohmic environment
for different temperatures $k_BT=$ 0 (full line), 0.005 (dotted line), 0.01
(short dashes) and 0.02 (long dashes) in units of $\hbar \protect\omega_R$.}
\end{figure}


\begin{references}
\bibitem{noise1}  For a review see Ya.M. Blanter and M. B\"uttiker,
Phys. Rep. {\bf 336}, 1 (2000).

\bibitem{exp-noise}  M.I. Reznikov et al. Phys. Rev. Lett. {\bf 75},
3340 (1995); H.E. van den Brom and J.M. van 
Ruitenbeek, Phys. Rev. Lett. {\bf 82}, 1526 (1999); R. Cron et al., 
submitted to Phys. Rev. Lett.

\bibitem{devoret}  M.H. Devoret et al., Phys. Rev. Lett. {\bf 64}, 1824
(1990); S.M. Girvin et al., Phys. Rev. Lett. {\bf 64}, 3183 (1990).

\bibitem{junctions} A.N. Cleland, J.M. Schmidt and J. Clarke, Phys.
Rev. Lett. 64, 1565 (1990); T. Holst, et al., Phys. Rev. Lett. {\bf 73},
3455 (1994).

\bibitem{cuevas}  J.C. Cuevas, A. Mart\'{\i}n-Rodero and A. Levy Yeyati,
Phys. Rev. B {\bf 54}, 7366 (1996).

\bibitem{keldysh}  L.V. Keldysh, Sov. Phys. JETP {\bf 20}, 1018 (1965).

\bibitem{comment}  It should be noticed that hopping events in the opposite
direction are connected by the correlator $P^{\alpha,\beta}$ while hopping
events in the same direction are connected by its inverse $%
1/P^{\alpha,\beta} $.

\bibitem{Khlus} V.A. Khlus, Sov. Phys. JETP {\bf 66}, 1243 (1987).

\bibitem{ingold} G.-L. Ingold and Yu.V. Nazarov in {\it Single Charge
Tunneling}, edited by H. Grabert and M.N. Devoret (Plenum Press, New York,
1992).

\bibitem{ruitenbeek}  J.M. van Ruitenbeek in {\it Mesoscopic electron
transport} Eds. L.L. Sohn, L.P. Kouwenhoven and G. Sh\"on, Kluwer Academic,
Dodrecht (1997).

\bibitem{scheer} E. Scheer et al., Phys. Rev. Lett. {\bf 78}, 3535 (1997) and
E. Scheer et al., Nature {\bf 394}, 154 (1998).

\bibitem{goffman}  M. Goffman et al., Phys Rev. Lett. {\bf 85}, 170 (2000).

\bibitem{zaikin} Coulomb blockade in a coherent conductor due to its own
resistance has been recently addressed by D.S. Golubev and A.D. Zaikin, 
cond-mat/0010493. For this situation both theories yield similar
results. 

\bibitem{weber} H.B. Weber et al., cond-mat/0007033.

\end{references}
\end{document}